# Quantum Cognition based on an Ambiguous Representation Derived from a Rough Set Approximation

Yukio-Pegio Gunji[1], Kohei Sonoda[2] and Vasileios Basios[3]

[1] *Department of Intermedia Art and Science, School of Fundamental Science and Technology, Waseda University, Ohkubo 3-4-1, Shinjuku, Tokyo, 169-8555 Japan*
[2] *Faculty of Education, Shiga University, Hiratsu 2-5-1, Otsu, Shiga, 520-0862 Japan*
[3] *Université Libre de Bruxelles, Statistical Physics and Complex Systems, Boulevard du Triomphe B-1050 Brussels, Belgium*

Over the last years, in a series papers by Arrechi and others, a model for the cognitive processes involved in decision making has been proposed and investigated. The key element of this model is the expression of apprehension and judgment, basic cognitive process of decision making, as an inverse Bayes inference classifying the information content of neuron spike trains. It has been shown that for successive plural stimuli this inference, equipped with basic non-algorithmic jumps, is affected by quantum-like characteristics. We show here that such a decision making process is related consistently with an ambiguous representation by an observer within a universe of discourse. In our work the ambiguous representation of an object or a stimuli is defined as a pair of maps from objects of a set to their representations, where these two maps are interrelated in a particular structure. The a priori and a posteriori hypotheses in Bayes inference are replaced by the upper and lower approximations, correspondingly, for the initial data sets that are derived with respect to each map. Upper and lower approximations herein are defined in the context of "rough set" analysis. The inverse Bayes inference is implemented by the lower approximations with respect to the one map and for the upper approximation with respect to the other map for a given data set. We show further that, due to the particular structural relation between the two maps, the logical structure of such combined approximations can only be expressed as an orthomodular lattice and therefore can be represented by a quantum rather than a Boolean logic. To our knowledge, this is the first investigation aiming to reveal the concrete logic structure of inverse Bayes inference in cognitive processes.

1. Introduction

Cognition in the process of decision making has two distinct features, namely *apprehension* and *judgment*. By apprehension we understand the coherent perception correlated with the recruitment and engagement in collective behavior of the required neuronal populations and by judgment we understand the process of comparing and of correlating -at least two- earlier apprehensions. Judgment requires memory and a certain degree of self-awareness, in the sense that the observer can hold on and internalize as its own a set of previously apprehended stimuli. Both processes require some kind of representation which entails their conceptualization; evidently both processes afford context dependencies within a certain framework of cognitive references. Arrechi critically discuss this issue in [1-5] and points out its quantum-like nature.

Recently theoretical and experimental advances in the area of quantum cognition and decision highlight this quantum-likeness of conceptualization. This endeavor brings together results from cognitive mathematical psychology [6], 'rational decision theory' and finance [7, 8, 9], experimental and theoretical artificial intelligence [10, 11], and quantum probability's logical structure [12, 13, 14]. Much is owned to the seminal work of Aerts from 1995 till now [15-24]. Nowadays, quantum-like behavior in language, concept correspondence and operational research and decision making have been developed in several publications by a considerable community of authors. For a comprehensive overview of the field one can consult the reference works covered in the recently published books by Busemeyer and Bruza [25], and by Khrennikov [26].

These authors focus on the probabilistic aspects and statistical verifications of structural aspects of quantum logic and quantum probability, while the studies of the dynamical basis of cognitive processes involved in decision making usually follow a more microscopic classical approach based on neural dynamics, see for example [25, 28]. Notably the role of chaotic dynamics in biological information processing, pioneered by the work of Nicolis and Tsuda [29] has been theoretically and experimentally investigated over the last decades by the group of Walter Freeman, K. Kaneko and others (an overview in [1, 29], particularly chapters 13, 15, 16, 17). Arrechi's proposal aims at bridging these two aspects, i.e. structural and dynamical, by constructing a model based on data sets from spike train synchronization and recursive (forward and inverse) Bayesian inference. The information content of the data sets of recorded neural spike-trains and analyzed by entropic measures enabling thus the tracing of suitable cross- and auto- correlation indexes. Based on such an approach Arrechi and his team conjecture and offer estimations for the existence of a quantal constant (Note: This is not the standard Plank constant of quantum physics but rather a quantum-like constant in the spirit of quantal neuron discharge proposed by Katz [31]) underlying specific quantum-like information processing for visual stimuli. As Arecchi demonstrates [32-34] the operation of the observing apparatus is equivalent to the operation of monitoring a Wigner distribution. The reported time coding implies a Wigner non-local measurement that stores and reads data in a global rather than a serial local fashion.

In the present work we built further on the procedure proposed by Arrechi and his team by considering an ambiguous representation based on "rough set analysis [35-41]" via a pair of mappings that serve as a model for Bayesian recurrent inferences within the conceptual categorization of stimuli afforded by an observer in a universe of discourse. We show that this can account, in a consistent way, for the quantum theoretical conjecture of Arrechi. We are able to show this by examining the underlying logic of the pair of mappings which turns out to be an orthomodular lattice as the ones related to quantum logic [42-44].

The paper is organized as follows, in Section 2 we briefly present the approach outlined by Arrechi and co-workers and discuss the inverse Bayesian procedure and its non-algorithmic aspects. Section 3 is the main exposition of our construction. In its first sub-section we develop the basic rationale for our rough set analysis framework. In the second sub-section we construct the two maps that implement an ambiguous representation based on the previous analysis and we offer specific examples for our model. In the third sub-section we discuss the derivation of a lattice of logical propositions from a rough set approximation. The fourth, and closing sub-section, of the third section presents the proof that the derived lattice of logical propositions coming from the above ambiguous representation is orthomodular, hence related to a quantum-logic lattice rather than a classical Boolean one. In Section 4 we revisit Arrechi's quantum-like hypothesis and we show that our scheme is consistent with his conjecture. Concluding in Section 5, we discuss the prospects of rough set approximation as a model for the inverse Bayes inference process and their non-algorithmic aspects and the outlook that this might entail for further investigations.

## 2. From apprehension to judgment and Quantum Bayes inference

In this section we briefly present the quantum theoretical conjecture proposed by Arecchi. As discussed in [2] *apprehension* consists of a coherent perception which emerges from the recruitment of neuronal groups, and *judgment* consists by recalling in memory previous apprehended units, coded in a suitable language, which subsequently are compared and selected in order to attain a judgment.

The first process, apprehension, has a duration around 1sec; its associated neuronal correlate can be traced by EEG (electroencephalogram) signals synchronized in the "gamma band" (frequencies between 40 and 60 Hz) coming from distant cortical areas. It can be described as an interpretation of the sensory stimuli on the basis of available algorithms, through a (forward) Bayes inference.

Following [1, 2, 3], let $h$ be the interpretative, competing, hypotheses in the presence of a sensory stimulus piece of data, $d$, the (forward) Bayes inference, selects the most plausible hypothesis, *h\**, that determines the motor reaction, exploiting an a priori existing algorithm, *P(d|h)*. That represents the conditional probability that a datum *d* is conforming with an hypothesis *h*. The *P(d|h)* are given e.g. have been learned during the past. They represent the equipment, or faculties, by which a cognitive agent, be it natural or artificial, faces and interprets its universe of discourse. For example, equipping a robot with a convenient set of *P(d|h)*, we expect a sensible behavior within a given environment providing sensory input translated to its repertoire of responses.

The second process, judgment, entails a comparison between two apprehensions acquired at different times, coded in a given language and recalled by memory. Let us call *d* the code of the second apprehension and *h\** the code of the first one, now -at variance with apprehension- it is *h\** that is already given; and instead the relation *P(d|h)* which connects them must be retrieved, it represents the conformity between *d* and *h\**, that is, the best interpretation of *d* in the light of *h\**. For example, as in [1, 2], two successive pieces of the text can be compared and the conformity of the second one with respect to the first one can be determined. This is very different from apprehension, where there is no problem of conformity but of plausibility of *h\** in view of a motor reaction. In [1, 2] the following two examples serve as illustration: "a rabbit perceives a rustle behind a hedge and it runs away, without investigating whether it was a fox or just a blow of wind. … As Walter Freeman puts it the life of a brainy animal consists of the recursive use of this inferential procedure [29]. … On the contrary, to catch the meaning of the 4-th verse of a poem, I must recover at least the 3-rd verse of that same poem, since I do not have a-priori algorithms to provide a satisfactory answer".

The bottom-up process entails the competition between two neuron groups fed by the same sensory stimulus *d*, while in the bottom-up procedure different interpretational stimuli, *P(d|h)*, provided by memory compete for selection. The one that prevails, as the corresponding top-down algorithm, *P(d|h)*, succeeds in better synchronizing neuron pulses of certain neural correlates than its concurrent competitors within the signal reader's "global workspace", where signals from different areas converge [45]. Thus, the winning hypothesis *h\** which will finally be driving the motor system is that which is provided by the selected neural correlate.

The whole process of top-down selection requires some form of self-awareness, and a representation scheme within the global workspace since the agent who performs the comparison must be aware that the two non-simultaneous apprehensions are submitted to its scrutiny. In contradistinction with apprehension, *judgment* does not presuppose a previously existing algorithm but it rather builds a new one, in an ad hoc fashion, through an ***inverse Bayes procedure*** [3]. It is exactly this construction of a new algorithm that Arrechi considers as the source of creativity and decisional freedom. In line with the pioneering investigations of Ernst Pöppel [46-48] the cognitive relevance of the 3sec interval for judgment has been also explored by Arrechi and his team on various experiments evolving subjects exposed to linguistic (literary and musical) texts.

Therefore, the process of judgment is not mechanical as it requires the ad hoc construction of a code for comparing and selecting two previously acquired apprehensions. Of course, once attained, judgments can be stored and memorized as resources for future reference. But each time a judgment is implemented will be a modified, even novel and hopefully enriched, version of the previous instances. Here is how Arrechi puts it [1] "As in any exposure to a text (literary, musical, figurative) a re-reading improves my understanding". The non-mechanical and non-algorithmic nature of the construction of a judgment is achieved by means of an ***inverse Bayes procedure*** [3, 4]. As far as the question concerning whether or not a computing machine can judge so that it can emulate a human cognitive agent (nb: referring to the Turing test), the answer is unequivocally "no". The condition that judgment entails "non-algorithmic jumps" is true as far as the *inverse* Bayesian inference process has to generate an *ad hoc* algorithm, build on the spot, data and context dependent, and by no means given beforehand.

Following the above line of reasoning Arrechi has formulated a quantum-like conjecture, aiming at explaining the necessary fast search within a semantic space; in a global rather than a local, sequential, exploration. This provides an uncertainty relation, ruled by a quantum-like constant, that yields a "decoherence" or de-correlation time compatible with the short term memory windows given by the experiments on linguistic understanding [1-4, 18-21]. In addition and to accord with this quantum-like line of investigation the "K-test" had also been considered [49], as the time equivalent of Bell inequalities in [5], and evidence is presented as a case of pseudo-violation of the Leggett-Garg inequality (p-LGI).

We further elaborate on the above exposition of apprehension and judgment by considering the role and position of an observer which is always a crucial issue in physics. As long as an observer stays outside of a universe of observations and retains a non-invasive role any objects within this "universe" can be described independently of the observer, and their properties can be defined without being influenced by the act of observation. By contrast, if an observer stays inside its "universe", we are faced with issues of interface and its interference on the designation of a cut-off between the inside, which is empirically known, and the outside which is empirically unknown [50-56]. In the case, where the act of observation disturbs the universe of observables an observer has to pay attention to (a) incompleteness and/or inconsistency in its universe which can be viewed as resulting from its own internal state perturbations, (b) chaos due to positive feedback loops between the act of observation and the properties of the objects observed, and (c) even stochasticity arising from constrains prohibiting complete determination of observables. An approach that aims to offer a description of a given universe viewed from an inside observer, has been proposed under the general term "endophysics" in [50] (for an easily accessible review by I. Tsuda and T. Ikegami, see [57]). The universe underpinning these internal perturbations due to the inclusion of the observer and the unavoidable self-referential issues could be described thanks to the chaotic dynamics in classical macroscopic systems [29, 30] or thanks to quantum mechanics in microscopic systems [58], but the problem of observer participation still remains as one of the most fascinating open problems [59].

As it is not the "quantum true matter" of quantum mechanics or the "Higgs ocean" or the "psi-waves" that we observe, but patterns of energy and information exchanged by the self and the environment through the various filters of the interfacing of our sensory organs, as W. Ebeling puts it, the deception has its roots in the evolution of life. In its particular form it differs significantly from organism to organism [60] (also in chapt. 9 in [1]). Each animal creates its "perceived objects" or its own cognitive space according to the attached survival relevance. Therefore, one of the greatest problems related to the issues of inside versus outside is the origin of self-consciousness and self-awareness or how can the brain be led to making self-consistent decisions against an unknown external world. In contemporary brain sciences the brain as a collection of neural cells is analyzed by their correlates, while single neurons are regarded as dynamical systems. The nonlinear nature of which can affected by chaotic dynamics, hence sometimes called as a chaotic neuron [29, 30]. The synchronization and correlation effects afforded by small groups of neuron in their chaotic regime are recently at the cutting edge of modern research [27, 28]. It is also customary that the perception can be viewed as the interplay of bottom-up stimuli processed by neurons locally and top-down interpretation excreting control in a global fashion. The fact that the whole process is affected by chaotic dynamics has been established both in theoretical and experimental grounds [61-63]. Thus the interplay of bottom up, top down processes is destined to be the generation of an array of probabilities of different decisions, and making a decision should account to the degeneration or "collapse" of this array of probabilities to one specific outcome. A particular motor decision emerges from the processing unit's global work space where the choice among competing signals arriving to it takes place. In some hypothesis, global work space acts as a threshold system by which the winner of competing neural groups is chosen [33-34, 45]. That is verified by evidence where the largest synchronized domain has occurred with each interpretation.

Apprehension can be described by successive applications of the (forward) Bayes inference using the same algorithm, and that corresponds to the analogy of the computation of climbing up an isolated peak by using only local information of the 'mountain' topology of the landscape. By contrast, the judgment corresponds to the analogy of the computation within climbing a 'mountainous landscape' with multiple peaks. Since a 'climber' uses only a local topological information such as the local gradients, one can reach not necessarily the global maximum but surely the local maximum, in this analogy the locally accessible highest peak. According to Arecchi, if a climber has an information on the global geometry, then the mixture of semantics (global geometry) and syntactical computation (choice of the step with the local biggest gradient) can entail an unreachable peak. Much like truths not decidable in Gödel's theorem of incompleteness [2]. How does the global "work space" [45] make a decision in judgment against this paradoxical situation? Non-algorithmic jump affected by the quantum-like effects could realize a judgment that corresponds to an inverse Bayes inference. While these ideas are examined by quantum Bayesian theory, a concrete mathematical model has not yet been proposed.

Independent of the quantum Bayesian approach, a class of "universes of discourse" inherent with ambiguity due to internal perturbations or lack of complete knowledge has been developed within the research field of soft computing by which ambiguous decisions could be allowed, in such approaches as in "fuzzy set"[64] and "rough set"[35-37] theories. Since the boundary of a fuzzy set or a rough set is indefinite and/or ambiguous, a small set of them can be used like an open set or neighborhood in the sense of specific topological spaces. Therefore, descriptions in a term of rough sets can be implemented as the logical structure of a partially ordered set or a lattice. This kind of ideas have been employed to the study of cognition with respect to discretized notions such as concepts, events and decisions. However, no study relating rough set approximation theory has been proposed for the issue regarding the inside versus outside role of the observer or in relation to quantum logic. To our knowledge, no such study within the various quantum Bayesian approaches appear to this effect. This is the main motivation which leads us to initiate our investigation for connections.

We show herein, using a technique based on rough set analysis, that the cut-off between the inside and outside universes of discourse of an observer can be expressed by an ambiguous representation for an object and its representation and that this ambiguous representation can be implemented by a pair of maps interrelated by a particular structure; moreover we show that the logic of such a pair of maps can be expressed as an orthomodular lattice like the ones familiar from quantum logic [12, 24, 25]. This demonstration accounts as a cognition model based on an ambiguous representation in terms of rough set lattice [39-41]. We also show that our model is consistent with the quantum approach proposed by Arecchi based on forward and inverse Bayesian inferences. A pair of rough set theory approximations for the data provided by an external stimulus is compared to the hypotheses, a priori and a posteriori respectively, while the inverse Bayesian inference can be expressed as a composition of two kinds of maps assigned to an ambiguous representation. Therefore, the above proposed scheme shows that the inverse Bayesian inference can be implemented by using our cognition model which is derived from rough set approximations.

## 3. Cognitive model based on ambiguous cognition

3-1 Inside vs outside duality implemented by the deconstructed axiom of choice

Human beings and other biological entities are faced with the outside surrounding themselves, they can perceive and recognize objects in the outside. How is their perception and/or cognition possible against the choice or cut off determining what is inside and what is outside? If this cognitive process is implemented or defined as a map such that objects in the outside are mapped into representations in the inside, the cut off of the inside vs outside is losing its sharp boundaries. Otherwise, if we focus on the crisp discontinuity between the inside and outside, how can such cognitive processes ever be implemented? Let us start by first introducing the discontinuity of the inside and outside of an observer by the difference between countable and uncountable sets.

Uncountable sets were first rigorously introduced and studied by Cantor via its famous diagonal argument. The work of Cantor on standard set theory implies that even if a set $X$, is equivalent to the set of natural numbers (i.e. countable), the power set of $X$ is uncountable. There is a cut-off between countable and uncountable sets such that a set in the power set of $X$ cannot be corresponded to an element in $X$ [65]. To overcome this cut-off, the "ZF axiomatic system" of set theory introduces a particular axiom called the "Axiom of Choice" which states that for any set there exists a "choice function". The axiom of choice is rewritten in modern category theory stating that any epimorphism has a section [66]. In other words, this implies that in set theory any onto map $f:A\rightarrow B$ has a particular map, section $r:B\rightarrow A$ such that $fr: B\rightarrow B$ coincide with an identity map of $B$. This is a condition under which a representative of a set can be chosen. Now, let us proceed by considering a set $A$ to be the 'outside' consisting of objects, and let $B$ be the 'inside' consisting of representations, and a map $f:A\rightarrow B$ to be a cognitive process relating outside objects to inside representations. We call $x$ and $f(x)$, an object and its representation, respectively. An equivalence relation $R$ on $A$ is derived by a map $f$, defined by $xRy$ for $x, y \in A$ such that $f(x)=f(y)$. It results in the 'outside' set of objects partitioned into a set of equivalence class such that

$$[x]_R = \{y \in A \mid xRy\}. \quad (1)$$

Due to the axiom of choice one can choose one representative $s$ for an equivalence class $[x]_R$. For example, imagine that the 'outside' consists of a three cats one tabby, one white, and one black two dogs, one a white dog and the other a black dog; and that the three cats as objects are mapped to their representative class CAT, and the two dogs are mapped to their representative class DOG. This results in two equivalence classes, {tabby cat, black cat, white cat} and {white dog, black dog}. Due to the axiom of choice one can choose a tabby cat as a representative for the class CAT and a white dog as a representative for the class DOG. Obviously it results to a one-to-one correspondence between the 'inside' and 'outside' if one neglects objects but not their representatives. In other words the axiom of choice implements a kind of 'illusionary' one-to-one correspondence between the inside and outside (Fig. 1). Therefore, such an isomorphism between the inside and outside can be achieved by the axiom of choice. The outside is, however, not trivial a priori. As objects are gradually and empirically identified as objects a posteriori, it suggests that is difficult for us to map objects to representations easily. The question we address now, as it readily arises, is how this situation can be implemented in a formal fraamework.

3-2 Ambiguity of representation implemented by two maps

We here implement "cognition against the cut-off between the inside and outside" by weakening the axiom of choice such that the choice of a representative is destined to be ambiguous. Or, as the above implies, that a representative cannot be uniquely chosen. Such an ambiguous choice is defined here by two maps $f$ and $g:A\rightarrow f(A)$. Let the equivalence relations derived from $f$ and $g$ be $R$ and $K$, respectively. Since a cognitive process $g$ conflicts $f$ for some $x \in A$ (i.e. $f(x) \neq g(x)$), the definition of $g$ has to be considered under a particular condition of an inhibition equipped with a symmetry and a locality law, as is discussed further below.

Before determining a representation of $x$ in a term of $g$, we introduce the inhibitory directed network $H$ defined by a quadruple $<E, V, d_s, d_t>$ where $E, V$ are the respectively the sets of edges and vertexes. And where $d_s:E\rightarrow V$ and $d_t:E\rightarrow V$ are the maps assigning a source and a target of an edge, respectively. We here define $V=f(A)$, and $d_s(E)=d_t(E)=f(A)$. Let the elements of $f(A)$ to be $b_1, b_2, \ldots, b_n$ (representation with respect to $f$). A representation (an element of $f(A)$) derives an equivalence class (a subset of $A$) which has no intersection with each other, therefore it is a partition. We here denote a partition derived by $b_k$ with respect to $f$ as

$$b_k^f = \{x \in A \mid f(x)=b_k\}. \quad (2)$$

Similarly an equivalence class with respect to $g:A\rightarrow f(A)$ is denoted by $b_k^g$.

The directed inhibitory network $H$ is sequentially determined under a symmetry law and locality law by the following: (i) Initialize $M$, a natural number by the cardinality of $f(A) \geq 4$, and $D$ a subset of $f(A)$ by $f(A)$: $M = |D|$. (ii) If $M > 0$, choose a natural number $m$ with $M-2 \geq m > 1$ randomly, where if $M$ is smaller than 4 then $m=M$, and otherwise ($0 \geq M$), the process is terminated. Then replace $M$ by $M-m$. (iii) Choose $m$ representations $y_1, y_2, ..., y_m$ in $D$ randomly and one element $x$ in $\{y_1, y_2, ..., y_m\}$ also randomly, and link $x$ with any $y_k$ by an edge $e_k$ such that $d_s(e_k)=x$ and $d_t(e_k)=y_k$ so as to satisfy a symmetry and locality law, however $b_k$ never link with $b_k$; Symmetry law requires that if there is an edge, $e$ such that $d_s(e)=b_p$ and $d_t(e)=b_q$ then there is an edge, $r$ such that $d_s(r)=b_q$ and $d_t(r)=b_p$; Locality law implies that if there are two edges $e$ such that $d_s(e)=b_p$ and $d_t(e)=b_q$ and $l$ such that $d_s(l)=b_p$ and $d_t(l)=b_u$ then there is an edge, $r$ such that $d_s(e)=b_q$ and $d_t(e)=b_u$. (iv) Elements chosen by the procedure (iii), $\{y_1, y_2, ..., y_m\}$, are eliminated from $D$. (v) The procedure of (ii) to (iv) is repeated till the procedure is terminated.

The symmetry law in the inhibitory network, for our example, implies that if an individual cat is not like a dog, an individual dog is not like a cat. Locality law implies that dissimilarity does not expand unnecessarily. Fig. 2 top left shows a part of an inhibitory network where $m=3$. Since the cardinality of $f(A)$ is 5, inhibitory network is not completed. Given a link from $b_2$ to $b_3$ and to $b_5$ in Fig. 2 top left ($m=3$), the symmetry law induces another links from $b_3$ to $b_2$ and from $b_5$ to $b_2$ (Fig. 2 top right) and the locality law induces a link from $b_3$ to $b_5$ (Fig. 2 middle left). Finally a part of inhibitory network is completed by a network among $b_2$, $b_3$ and $b_5$ only due to the symmetry and locality law (Fig. 2 middle right). When a source $b_1$ links with a target $b_4$ and a source $b_4$ links with a target $b_1$ as shown in Fig. 2 bottom left, an inhibitory network is completed. Inhibitory network is represented by a matrix as shown in Fig. 2 bottom right. Take an element of a matrix at a column $b_j$ and a row $b_i$. If there is a link between $b_i$ and $b_j$, in the inhibitory network, then a component at ($b_i$, $b_j$) is blank (0), and otherwise it is painted in black (1) in a matrix. The obtained matrix consists of a diagonal matrix whose diagonal elements are 1 and others are 0's, with a background whose components are all 1's.

Directed inhibitory network $H$ straightforwardly determines a secondary map $g:A \rightarrow f(A)$. Imagine that there are satisfactory many elements in each equivalence classes with respect to $f$. Referring to the network $H$ as shown in Fig. 2, an element in an equivalence class is mapped to an element of $f(A)$ as shown in Fig. 3. Elements in an equivalence class, $b_1^f$ are mapped to any elements of $f(A)$ except for $b_4$ because a matrix in Fig. 2 bottom right shows that $b_4$ in $g(A)$ (column) is linked with $b_1$ in $f(A)$ (row) in the inhibitory network $H$. Similarly elements in $b_2^f$ is mapped to any elements of $f(A)$ except for $b_3$ and $b_5$ due to the inhibitory network and then they are mapped to $b_1$, $b_4$ and $b_2$.

If the secondary map $g$ is defined so as to satisfy the directed inhibitory network, the relation between equivalence classes (i.e. partitions) with respect to $f$ and $g$ can be rearranged to a particular matrix which consists of various size of diagonal matrices. Fig. 4 top left shows a given matrix produced by a relation between equivalence classes with respect to $f$ and $g$, where if there exists an element $x$ in $A$ $f(x)=i$ and $g(x)=j$ then a component at ($i$, $j$) is 1 and otherwise 0. The algorithm to construct a matrix consisting of diagonal matrices is started from collecting a column which has 0 component at the same row. The 4$^{th}$ components in the 1- and 3-column in a given matrix are 0, and the 1$^{st}$ components in the 1- and 6-colums are 0. Then all columns, 1-, 3- and 6-columns are pasted with each other, and are constructed to a block. Similarly a block consisting of 2-, 4- and 5-columns is formed and a similar block is obtained. As the obtained two blocks are formed, a matrix rearranged in a term of columns is further obtained (Fig. 4 bottom).

Fig. 5 shows a similar procedure in a term of rows. Finally we obtain a matrix consisting of diagonal matrices equipped with 1-components background as shown in Fig. 5 bottom left. Because columns having 0 component at the same row, and rows having 0 component at the same column can be arranged in the form of a diagonal matrix, it can be straightforwardly verified that this algorithm can lead to a matrix consisting of diagonal matrices.

Fig. 6 shows a matrix consisting of diagonal matrices computed by this procedure which are constructed for a given relation between equivalence classes with respect to $f$ and $g$ satisfying the inhibitory relation, where component of 1 and 0 are reversely represented, compared to the matrix shown in Fig. 4 and 5. If there is no element $x$ in $A$ such that $f(x)=i$ and $g(x)=j$, then a component at

(*i, j*) is 0 (blank), otherwise it is 1 (black). An matrix obtained following this shows a 'universe' consisting of 'sub-universes' for which any concept in that 'sub-universe' can be constructed by a possible union of equivalence classes. However, the sub-universes are partially overlapping each other, and therefore some concepts which would be possibly constructed by union of equivalence classes employed to different sub-universes cannot be obtained. We turn now to the question of characterization of the whole 'universe' in terms of its logic structure. Using rough set analysis [41] we subsequently show that the structure of the whole 'universe' can be verified to be an orthomodular lattice in terms of lattice theory [42-43, 67-68].

3-3 A lattice derived from a rough set approximation

As mentioned in section 3-1, a map $f:A \rightarrow B$ induces an equivalence relation $R$ such that $xRy \Leftrightarrow f(x)=f(y)$. By using such an equivalence class of $R$, any subset $X$ in $A$ can be approximated as a rough set in upper or lower approximation. The upper approximation of $X$ with respect to $R$, $R^*(X)$ is defined by

$$R^*(X) = \{y \in [x]_R \mid [x]_R \cap X \neq \emptyset\}. \tag{3}$$

If an equivalence class is compared to a neighborhood in topology, upper approximation is analogous to closure. Lower approximation of $X$ with respect to $R$, $R_*(X)$ is defined by

$$R_*(X) = \{y \in [x]_R \mid [x]_R \subseteq X\}. \tag{4}$$

The lower approximation is analogous to the set of the interior points in topology. It is clear to see that for any $x \in A$

$$R^*([x]_R)=[x]_R, \quad R_*([x]_R)=[x]_R. \tag{5}$$

Thus any union of equivalence classes is also a fixed point with respect to $R^*$ and $R_*$. We define a lattice $<L_R, \subseteq>$ such that $L_R$ is a collection of fixed points satisfying

$$R^*(X)=X \tag{6}$$

for $X \subseteq A$ and they are ordered by inclusion. A lattice is a partially ordered set closed with respect to meet and join operations of two elements, where the join of two elements is the least element which is larger than both of two elements and the meet of two elements is the greatest element which is smaller than both of two elements [67, 68]. The meet in $L_R$ is defined by intersection, and the join is defined by union, which implies $L_R$ is a set lattice and is a Boolean algebra. In other words, any of the concepts in a set lattice can be constructed by possible combinations of atoms which are partitions. If a fixed point (6) is replaced by $R_*(X)=X$, then the same Boolean algebra is obtained.

How is a set lattice $<L_R, \subseteq>$ represented in the form of a matrix between partitions? Because a partition has no intersection with each other, only a diagonal component is 1 (i.e. $f(b_i)= f(b_i)$) in a matrix whose rows and columns are partitions of $R$. In other words, $<L_R, \subseteq>$ is represented by a matrix whose diagonal components are 1's and any other components are 0's of the relation between partitions, which is nothing but a diagonal matrix. In other words, a set lattice or a Boolean algebra is represented by a diagonal matrix.

3-4 Orthomodular lattice derived from ambiguous representation

We now proceed in characterizing the lattice that represents the relation between partitions of *f* and *g* satisfying the inhibitory network (i.e. a matrix consisting of diagonal matrices). This can be readily illustrated in terms of a concrete implementation, for example by using a diagonal matrix shown in Fig. 5bottom left, we can estimate the properties of its lattice. Number the column and rows in a given matrix, from the top to the bottom, or from the left to the right, $b_1, b_2, \ldots, b_6$, respectively. Assume that rows represent partitions derived by $g$, and columns represent partitions derived by $f$. In this sense a component which is 1 at $(i, j)$ shows that there exists an element $x$ in $A$ such that $f(x) = g(x)$, or that $b_i^g \cap b_j^f \neq \emptyset$. Call the above left diagonal matrix, $B_1$, and the below right one, $B_2$. Since a matrix implies ambiguity of representation, $f$ and $g$, the approximation in a term of a rough set has to compose two kinds of equivalence relation, $R$ derived by $f$, and $K$ derived by $g$. Fixed points with respect to the composition of two kinds of rough set approximation are collected in order to obtain a lattice. Let us now demonstrate how this composition is obtained and its lattice properties:

Assume that the composition of approximation is defined by the order from $K$ to $R$. Under this assumption, we can consider four kinds of compositions, $R^*K^*$, $R_*K_*$, $R_*K^*$ and $R^*K_*$. We here estimate these compositions with respect to whether a partition is allowed as a fixed point. Consider the case of $R^*K^*$. We readily obtain that $R^*K^*(b_1^f) = R^*(\{y \in [x]_K \mid [x]_K \cap b_1^f \neq \emptyset\}) = R^*(b_1^g \cup b_4^g \cup b_5^g \cup b_6^g) = b_1^f \cup b_2^f \cup b_3^f \cup b_4^f \cup b_5^f \cup b_6^f = A$. This implies that a partition $b_1^f$ is not a fixed point with respect to $R^*K^*$. Next we consider the case, $R_*K_*$. Actually, $R_*K_*(b_1^f) = R_*(\{y \in [x]_K \mid [x]_K \subseteq b_1^f\}) = R_*(\emptyset) = \emptyset$. Therefore it follows that $b_1^f$ is not a fixed point with respect to $R_*K_*$. Next we consider the case of the $R_*K^*$ composition, we see that $R_*K^*(b_1^f) = R_*(\{y \in [x]_K \mid [x]_K \cap b_1^f \neq \emptyset\}) = R_*(b_1^g \cup b_4^g \cup b_5^g \cup b_6^g) = b_1^f$. This verification can be generalized to any number of partitions. Thus we can say that any partition with respect to $f$ can be a fixed point $X$ for the equation,

$$R_*K^*(X) = X. \tag{7a}$$

Finally for the composition $R^*K_*(X)$ we observe that, as we previously showed, a collection of a solution for equation (5) is a lattice whose partial order is defined by inclusion, and that lattices defined by a collection of fixed points for $R_*K^*(X) = X$, $R^*K_*(X) = X$, $K_*R^*(X) = X$ and $K^*R_*(X) = X$ are isomorphic to one another. Consequently we can estimate the logic of a universe consisting of ambiguous representation, by a fixed point with respect to an operation, $R_*K^*$ or $R^*K_*$. It implies that a fixed point

$$R^*K_*(X) = X \tag{7b}$$

is also adopted to construct a lattice.

Next we estimate the detail of logical structure of a collection of fixed points mentioned above. In referring to Fig. 5 bottom left, we obtain $R_*K^*(b_1^f \cup b_2^f) = R_*(\{y \in [x]_K \mid [x]_K \cap (b_1^f \cup b_2^f) \neq \emptyset\}) = R_*(b_1^g \cup b_2^g \cup b_4^g \cup b_5^g \cup b_6^g) = b_1^f \cup b_2^f$. Similarly, $R_*K^*(b_1^f \cup b_3^f) = b_1^f \cup b_3^f$, and $R_*K^*(b_2^f \cup b_3^f) = b_2^f \cup b_3^f$. However, $R_*K^*(b_1^f \cup b_2^f \cup b_3^f) = R_*(A) = A$. It is clear to see that the least element of a lattice, an empty set is a fixed point with respect to any operators of rough set approximation, and that $R_*K^*(\emptyset) = \emptyset$. Thus the block $B_1$ reveals all possible union of partitions except for a union of all partitions constituting $B_1$. The block $B_2$ also satisfies the same condition. Therefore, we obtain all elements of a lattice obtained from a matrix as shown in Fig. 5 bottom left, by $\{\emptyset\} \cup \{b_1^f, b_2^f, b_3^f, b_1^f \cup b_2^f, b_2^f \cup b_3^f, b_1^f \cup b_3^f\} \cup \{b_4^f, b_5^f, b_6^f, b_4^f \cup b_5^f, b_5^f \cup b_6^f, b_4^f \cup b_6^f\} \cup \{A\}$.

This construction can be generalized for any matrix consisting of diagonal matrices with a background of 1 components. Assume that a matrix consists of a block, $B_1, B_2, \ldots, B_n$, where each block, $B_k$ consists of 1-component $(b_{k1}, b_{k1}), (b_{k2}, b_{k2}), \ldots, (b_{ks}, b_{ks})$. Simply, we state that $B_k$ consists of $b_{k1}, b_{k2}, \ldots, b_{ks}$. Each block yields elements of a lattice such as

$$\bigcup_{i=1}^{I(u,w)} b_{kp(w,i)}^f = b_{kp(w,1)}^f \cup b_{kp(w,2)}^f \cup \ldots \cup b_{kp(w,I(u,w))}^f \tag{8}$$

where $p(w, i) \in \{1, 2, \ldots, s\}$ with $p(w, i) \neq p(w, j)$ for $i \neq j$, and $I(u, w) = 2^{u-1}$ for $u = 1, 2, \ldots, s-1$, and

$w=1, \ldots, {}_sC_u$ for each $u$, and where $\cup^{I(u,w)} b_{kp(w,i)}{}^f \neq \cup^{I(u,v)} b_{kp(v,i)}{}^f$ for $w \neq v$. Consider the case with $s=3$, of which $B_k$ consists of $b_{k1}$, $b_{k2}$ and $b_{k3}$. If $u=1$, then $w=1, 2, 3$, $I(1, 1)=I(1, 2)=I(1, 3)=2^0=1$, and elements of a lattice, $\cup^{I(1, 1)} b_{kp(1, i)}{}^f = b_{kp(1, 1)}{}^f = b_{k1}{}^f$, $\cup^{I(1, 2)} b_{kp(2, i)}{}^f = b_{kp(2, 1)}{}^f = b_{k2}{}^f$, $\cup^{I(1, 3)} b_{kp(3, i)}{}^f = b_{kp(3, 1)}{}^f = b_{k3}{}^f$. Since $b_{kp(1, 1)}{}^f \neq b_{kp(2, 1)}{}^f$, $b_{kp(2, 1)}{}^f \neq b_{kp(3, 1)}{}^f$, three different elements can be obtained. Similarly, if $u=2$, then $I(2, 1)=\ldots=I(2, {}_3C_2)=I(2, 3)=2^1=2$ and elements of a lattice, $\cup^{I(2,1)} b_{kp(1, i)}{}^f = b_{k1}{}^f \cup b_{k2}{}^f$, $\cup^{I(2,2)} b_{kp(2, i)}{}^f = b_{k1}{}^f \cup b_{k3}{}^f$, $\ldots$, $\cup^{I(2,3)} b_{kp(3, i)}{}^f = b_{k1}{}^f \cup b_{k3}{}^f$. Since $u \neq s$, then a union of all partitions of $b_{k1}, b_{k2}, \ldots$ and $b_{ks}$ is not an element of a lattice. Instead, the least and the greatest element of a lattice, which are

$$A, \quad \varnothing, \tag{9}$$

respectively, are also elements of a lattice. It is easily verified that $R_*K^*(A) = A$. We call a lattice whose elements are defined by (8) and (9), a block lattice $L(B_k)$, where a block lattice is derived by a block matrix $B_k$. It is clear to see that each $\cup^{I(1, w)} b_{kp(w, i)}{}^f = b_{kj}{}^f$ is an atom of a block lattice, where an element $x$ of a lattice is an atom if and only if there is no any other element between $x$ and the least element of a lattice.

Further we show that the union of atoms which belong to different block lattices is not a fixed point with respect to $R_*K^*$: Let different block lattices in a given lattice derived from diagonal matrices, $L(B_k)$ and $L(B_h)$, and take an atom $b_{ki}{}^f$ from $L(B_k)$ and an atom $b_{hj}{}^f$ from $L(B_h)$. It is straightforward to see that $R_*K^*(b_{ki}{}^f \cup b_{hj}{}^f) = R_*(A) = A$, and that $b_{ki}{}^f \cup b_{hj}{}^f$ is not a fixed point of $R_*K^*$. Because background of diagonal matrices consists of 1-components, $A - b_{(k)}{}^g$ is related to $b_{ki}{}^f$ and $A - b_{(h)}{}^g$ is related to $b_{hj}{}^f$, where $b_{(k)}{}^g = \cup b_{kp}{}^g$ and $p$ can be $1, 2, \ldots$ except for $i$, and $b_{(h)}{}^g = \cup b_{hq}{}^g$ and $q$ can be $1, 2, \ldots$ except for $j$. Thus,

$$K^*(b_{ki}{}^f \cup b_{hj}{}^f) = (A - b_{(k)}{}^g) \cup (A - b_{(h)}{}^g) = (b_{(k)}{}^g)^c \cup (b_{(h)}{}^g)^c = (b_{(k)}{}^g \cap b_{(h)}{}^g)^c = \varnothing^c = A, \tag{10}$$

where $B^c$ represents a complement of $B$ such that $B \cap B^c = \varnothing$ and $B \cup B^c = A$. It shows that union of atoms from different block lattices is not a fixed point respect to $R_*K^*$ and it is not an element of a lattice for a given matrix. It also means that there is no element overlapping in different block lattices, except for the least and the greatest element.

Each block lattice is a Boolean algebra. Each block lattice consists of elements of a set lattice except for a union of all partitions, and the greatest element of a whole lattice, $A$ which can be also the greatest element of each block. That is why each block lattice, $L(B_k)$ from which a union of all partitions is removed and to which $A$ is added, is isomorphic to a set lattice that is a Boolean algebra. Consequently, we can say that a lattice derived from a matrix which consists of $n$ diagonal matrices with a background of 1-components, $L_D$ is expressed as

$$L_D = \bigcup_{k=1}^{n} L(B_k). \tag{11}$$

Each $L(B_k)$ is a Boolean algebra. For any $i, j \in \{1, 2, \ldots, n\}$ with $i \neq j$,

$$L(B_i) \cap L(B_j) = \{A, \varnothing\}. \tag{12}$$

The above shows that $L_D$ is an almost disjoint union of Boolean algebras. To estimate whether a lattice $L_D$ is an orthomodular lattice or not, we have to refer to the presence of $n$-rank loop in a lattice, which is defined by the following: A finite series of Boolean algebras, $L(B_1), L(B_2), \ldots, L(B_n)$ is a loop with rank $n$ if this series satisfies

$$L(B_i) \cap L(B_{i+1}) = \{A, \varnothing, Z, Z^c\} \tag{13a}$$

$$j \notin \{i\text{-}1, i, i\text{+}1\}, \quad L(B_i) \cap L(B_j) = \{A, \emptyset\} \tag{13b}$$
$$i, j, k \in \{1, 2, \ldots, n\}, \quad i \neq j \neq k, \quad i \neq k, \quad L(B_i) \cap L(B_j) \cap L(B_k) = \{A, \emptyset\}. \tag{13c}$$

In equation (13a) $Z$ represents an atom common in both $B_i$ and $B_{i+1}$. An orthomodular lattice is defined by an almost disjoint union of Boolean algebras without a loop with rank 3 or 4. The $L_D$ never contains a pair of block satisfying equation (13a) because of (12), $L_D$ never contains a loop of rank 3 or 4. Especially if $L_D$ contains no loop, it is an orthocomplemented lattice. It shows that $L_D$ is an orthomodular lattice, and it is also an orthocomplemented lattice.

Fig. 7 shows a lattice derived from a matrix shown in Fig. 5 bottom left. Since a matrix consists of two diagonal blocks, a lattice is a quasi-disjoint union of two Boolean algebras. Each diagonal matrix contains three diagonal elements, and that shows $2^3$-Boolean algebra. Each element of a lattice is represented by a circle in Hasse diagram as shown in Fig. 7. If $X \subseteq Y$, an element $X$ is located below an element $Y$, and if there is no element between $X$ and $Y$, two elements are linked by a line. Each lattice consisting of eight elements represents $2^3$-Boolean algebra. Since the greatest ($A$) and the least element ($\emptyset$) are common to two $2^3$-Boolean algebras, a lattice corresponding to a matrix in Fig. 5 is represented by a Hasse diagram in Fig. 7 right. It implies that a lattice represents an orthocomplemented lattice. Thus, it is also an orthomodular lattice. Similarly, Fig. 8 shows a Hasse diagram of a lattice corresponding to a matrix shown in Fig. 6 above. Since a matrix consists of a diagonal matrix containing six elements, two diagonal matrices with five elements or two elements, a lattice consists of one $2^6$-Boolean algebras, two $2^5$-Boolean algebras and two $2^2$-Boolean algebras. Broken lines are used to indicate that the greatest and least elements represented by circles are common to all block lattices (they are not used to indicate inclusion relation or order). The above testifies also that the lattice at hand is an orthomodular (and orthocomplemented) lattice.

According to Birkhoff and Neumann [69], the closed subspace of a Hilbert space formulation of quantum mechanics forms an orthomodular lattice. Although Mackey attempted to provide an axiomatic system of quantum logic as an orthocomplemented lattice, it is has not been satisfactorily demonstrated [13]. An orthomodular lattice is of crucial importance to quantum logic [42, 43]. Our argument shows that human cognition appearing in as a macroscopic phenomenon also reveals an inherent quantum logic and therefore quantum-like effects. How is our model related to quantum conjectures in cognitive systems is the subject of the next section.

## 4. Quantum Bayes inference

In this section we show that our work is consistent with quantum theoretical conjecture proposed by Arecchi and discussed in Section 1. As mentioned before, decision to an external stimulus could result from a choice of a neural group which can give rise to largest synchronized and distributed domains of correlates. Such a process called apprehension can be implemented by (forward) Bayes inference as proposed by Arecchi:

$$P(h^*) = P(h|d) = P(h)P(d|h)/P(d), \tag{14}$$

where $P(h)$ is an a priori probability of hypothesis $h$, $d$ is data, $P(d|h)$ is a priori probability that $d$ results from $h$, $P(d)$ is a probability we observe data $d$, and $P(h^*)$ is a posteriori probability among a priori hypothesis. Since Bayesian inference is compared to a process of climbing a mountain within a landscape of coded data searching by using only local information, $P(h^*)$ provides the next plausible climbing step obtained from the previous step, $P(h)$, provided by the data from an external stimulus, $P(d)$, and especially of the external stimulus received at the position of the previous step, $P(d|h)$. If an isolated unique stimulus is given, the process of data apprehension occurs.

When two successive stimuli are given within a specified short term, the first apprehension stored is retrieved and is compared to the second one. Thus it can give rise to the process of judgment. If this process is also to be compared to the climbing of a mountain within a landscape of coded data, the mountain would contain multiple peaks which prohibits a climber only with local view to climb to the globally highest peak. Globally optimum solution is obtained not by logical computation but by a non-logical jump. According to Arecchi, the non-logical jumps appearing in cognition provide evidence for quantum-like effects, and this is implemented by an inverse Bayes inference. From equation (14), the inverse Bayes inference is expressed by

$$P(d|h) = P(d) P(h^*) / P(h). \tag{15}$$

Where now the data (*d*) and a posteriori hypothesis (*h\**) represent the current stimulus and the previous one, respectively. The most suitable algorithm, $P(d|h)$ that best matches *d* and *h\** is obtained by an inverse Bayes inference. Since the inverse Bayes inference contains non-algorithmic jump, $P(d|h)$ is not a simple modification of the formula from equation (14) but an equation obtained from modification of the hypothesis, $P(h)$. Therefore, $P(d|h)$ cannot be obtained by simple iterations of recursive process, hence its non-algorithmic nature.

Quantum Bayes inference proposed by Arecchi can be compared to our cognitive model based on ambiguous representation. First we replace terms in Bayes inference by our data and approximations. Because a given set of data *d* in Bayes inference can be replaced with a given subset *X* in a universal set *A*, and because lower and upper approximations for *X* have the following order,

$$R_*(X) \subseteq X \subseteq R^*(X), \tag{16}$$

data coupled with a priori hypothesis such as $P(d|h)$ can be replaced by the upper approximation, $R^*(X)$, and data coupled with a posteriori hypothesis such as $P(h|d) = P(h^*)$ can be replaced by the lower approximation, $R_*(X)$. From inequality (16), any element in the lower approximation is *X*, and only some elements in the upper approximation are elements of *X*. Thus, the lower and upper approximation could correspond to satisfactory and necessary condition, respectively. That is why the upper and lower approximation could be considered to correspond to the a priori and a posteriori hypotheses, respectively.

How is the Bayesian inference implemented in our ambiguous representation? In equation (14), *h* is replaced by equivalence relation, *R* derived by a particular representation (map), $P(d|h)$ is replaced by $R^*(X)$, and $P(h|d) = P(h^*)$ is replaced by $R_*(X)$. Therefore, Bayesian inference is the process of computing $R_*(X)$ from $R^*(X)$ (i.e. from a priori to a posteriori). So as to achieve this computation data, *X*, is implemented by a particular form because $R^*$ is used instead of an equivalence class, $[x]_R$. Finally the Bayesian inference is expressed, for any subset *X* in a universal set *A*, by

$$R_*(X) = \{y \in R^*(\{x\}) \mid R^*(\{x\}) \subseteq X\}. \tag{17}$$

It is clear to see that equation (17) can be verified because $R^*(\{x\})=[x]_R$. In this sense the inverse Bayes inference is the process of computing $R^*(X)$ from $R_*(X)$ (i.e. from a posteriori to a priori). This process however, does not hold as

$$R^*(X) \neq \{y \in R_*(\{x\}) \mid R_*(\{x\}) \cap X \neq \emptyset\}. \tag{18}$$

If $[x]_R \supset \{x\}$, $R_*(\{x\})=\emptyset$, and then even if $R^*(X)$ is not empty, the right hand term in equation (18) is empty if for all *x* in *X* $[x]_R \supset \{x\}$.

The question that arises now is how the inequality in equation (18) can be replaced by equality. We can answer this question by finding a particular condition under which equality in (18) holds. Since *X* is any subset of a universal set, there exists a case of which $X=\{x\}$. We generalize such a case and replace $\{x\}$ and *X* by a particular set *X* that is a subset of the universal set, and obtain

$$R^*(X) \sim \{y \in R_*(X) \mid R_*(X) \cap X \neq \emptyset\}. \tag{19}$$

The symbol ~ just signifies a relation which contains a particular equivalence relation as a subset. Because equivalence relation $R$ never equate the left with right term in (19), an equivalence relation has to be replaced by another one, $K$ which can be expected to equate the left with the right term for more subsets $X$ in the universal set $A$. Compared to the inverse Bayesian inference, to compute an a priori hypothesis $R^*(Y)$, a non-algorithmic jump related to quantum-like effects corresponds to replacing a relation $R$ by another relation $K$. We also implement the way to collect an element, $y \in R_*(X)$ by $y \in Y$, where $Y$ is an equivalence class with respect to $R$, and restrict $Y$ to a subset $X$. Then,

$$R^*(Y) \sim \{y \in Y \mid K_*(Y) \cap Y \neq \varnothing\}. \tag{20}$$

Since $Y$ is an equivalence class with respect to $R$, $Y$ per se can play a role in collecting an element. The right hand term in relation (20) implies the upper approximation of $K_*(Y)$ with respect to $R$ insofar as $Y$ is an equivalence class with respect to $R$. Since $K_*(Y) \subseteq Y$, if $K_*(Y) = \varnothing$, $\{y \in Y \mid K_*(Y) \cap Y \neq \varnothing\} = \varnothing$, and otherwise $\{y \in Y \mid K_*(Y) \cap Y \neq \varnothing\} = Y$. Then we obtain

$$R^*(Y) = \{y \in Y \mid K_*(Y) \cap Y \neq \varnothing\} = R^*(K_*(Y)), \tag{21}$$

if $K_*(Y) \neq \varnothing$. This implies that there is a particular relation between equivalence classes with respect to $K$ and $R$ including a condition such that $K_*(Y) = [x]_K$ and $R^*([x]_K) = Y$. That is a condition which is satisfied by a pair of equivalence relations derived by a pair of maps, $f$ and $g$ under a constraint of an inhibitory network equipped with symmetry and locality. Such a pair of equivalence relations can constitute a matrix consisting of diagonal matrices with 1-components background. It is also crucial in comparing our model to that of Arecchi's that a collection of fixed points is naturally introduced in the form of (21), $R^*(K_*(Y)) = Y$.

Finally we conclude that the inverse Bayesian inference can be implemented by equation (21) which contains a non-logical jump such as the modification of the original hypothesis (i.e. a novel equivalence relation) from $R$ to $K$. Especially, since there is a particular relation between $R$ and $K$ of which a relation between them constitutes a matrix consisting of diagonal matrices with 1-components background, a lattice resulting from a collection of fixed points satisfying $R^*(K_*(Y)) = Y$ is an orthomodular lattice. In other words, the non-logical jump is contained within the inverse Bayesian inference and could refer to quantum-like effects. Therefore our cognitive model is consistent with the inverse Bayesian inference suggested by Arecchi.

## 5. Conclusion

This work aims in contributing to the discussion of apprehension and judgment as processes studied within the scope of self-awareness or self-consciousness as studied in terms of neural correlations. As certain aspects of brain-machine analogy are progressing, our work can give rise to the implementation of algorithmic procedures aiding to decision making, such as the Bayesian inference, in its classical and quantum versions. It also strengthens analogies in thought with respect to functions and faculties of consciousness as integrated information representation and processing. As brain sciences search for particular network structures by which information is integrated and embedded our approach can be regarded as aiming towards understanding the basic features of the globally superimposed constraints networks through macroscopic correlations giving rise to judgment and even conscious discrimination.

Regardless of the progress of brain-machine analogy, decision making related to a successive apprehension containing judgment could appeal towards the limits of this analogy in a way to a brain-machine dis-analogy. Because the judgment cannot be implemented by a computation in which the same algorithm is successively applied to data, and has to be equipped with non-algorithmic jump affected by the quantum-like effects and an underlying quantum-like logic. It can also appeal that judgment, conceptualization and/or category creation in human cognition has to be non-algorithmic processes, and that they cannot be implemented by a machine in a strict algorithmic fashion.

Crowds of cortical neurons provide the appropriate same substratum, in terms of complexity flexibility, adaptability and plasticity, for collective agreement and synchronization in both instances of apprehension and judgment. A fascinating question is whether or not this is the only such substratum in existence. Research in collective decision making and event anticipation in other crowds -that of social animals such as bees, ants, even *physarum polycephalum* and other model systems [70]- reveal certain analogies in recruiting and consensus building. Our approach might help in shedding new light and in instigating research towards this kind of investigations.

The inverse Bayesian inference containing non-algorithmic jumps or referring to quantum-like aspects is here supported and related to the process of approximation with respect to rough set analysis. Rough set analysis here captures the minimal requirements of a coarse grained representation of the environment through the sensory and conceptual filters of the observer. Ebelling and Feisel (see chap 9 of [1] and also [60]) ascribe to information processing the key role to survival of any living observer. In the abstract framework chosen we are able to capture the minimal aspects of these fundamentally self-referential feedback loops and its primary semantics while probing the environment, "out there", at the instances of apprehension and judgment, "from the inside".

The formation of hypotheses per se can correspond to the equivalence relation derived from the mapping from objects to their representations. The a priori and a posteriori hypotheses in Bayesian inference were shown to correspond to the upper and lower approximations for an object in rough set analysis. While Bayesian inference was expressed as a formulation of the lower approximation based on the upper approximation (i.e. degeneracy of hypothesis a posteriori from a priori), the inverse Bayesian inference has to be implemented by a formulation of the upper approximation based on the lower approximation with modification or replacement of the algorithms in question. It is, therefore, implemented by composition of different approximations with respect to different equivalence relations. That is nothing but a non-algorithmic jump revealing the associated quantum-like effects reported in the literature. The approximation proposed, based on the composition of two kinds of equivalence relations, leads to a logic expressed as an orthomodular lattice. Conversely it reveals that the modification or replacement of algorithms (i.e. replacement of equivalence relations) is affected by the reported quantum-like effects.

Even if it is assumed that neurons constituting a brain are expressed as chaotic neurons, brain-machine dis-analogy referring to the non-algorithmic jump can be found. The non-algorithmic jump can be further investigated as the non-algorithmic interaction between syntax and semantics through a particular mixture or ambiguity open to both syntax and semantics or equivalently to both local and global properties. We demonstrated here, thanks to a formulation of the inverse Bayesian inference, that computing processes expressed as rough set approximations with non-algorithmic jumps are expressions of the orthomodularity of the underlying lattice which refers to a quantum logic or to quantum-like effects in cognition. Furthermore, our work suggests that complex and ambiguous macroscopic phenomena such as cognition and decision making can be related to quantum theory at large.

## Acknowledgements

Vasileios Basios wishes to thank the NSRF "Thales" program for support. This research has been co-financed by the European Union (European Social Fund -ESF)and Greek national funds through the Operational Program "Education and Lifelong Learning" of the National Strategic Reference Framework (NSRF) -Research Funding Program: Thales. Investing in knowledge society through the European Social Fund.

# Figure captions

Figure 1. Schematic diagram showing the one-to-one corresponding between objects and representations due to the axiom of choice. A map, $f$ from the outside (objects) to the inside (representations) entails partitions in the outside (left above). Thanks to the axiom of choice, one representative can be chosen for each partition (left below). In ignoring all elements but representatives, one can find the one-to-one correspondence between the 'inside' and 'outside' (right).

Figure 2. Construction for directed inhibitory network of $<E, V, d_s, d_t>$ where $V=\{b_1, b_2, b_3, b_4, b_5\}$. Initially one vertex is chosen randomly and edges whose source is a chosen vertex are also randomly chosen (top left). By using symmetry (top right) and locality law (middle left) a local network is obtained (middle right). Successive applying the procedure for all elements of $V$ leads to a directed inhibitory network (left bottom) which can be expressed as a particular matrix (right bottom).

Figure 3. Construction for a pair of maps, $f$ and $g$ from a set of objects, $A$ to a set of representations $f(A)$. Given a map $f:A \rightarrow B$, the other map $g$ is determined so as to satisfy the directed inhibitory network.

Figure 4. Construction for a matrix consisting of diagonal matrices with background of 1-components, with respect to rearrangement of columns.

Figure 5. Construction for a matrix consisting of diagonal matrices with background of 1-components, with respect to rearrangement of rows. It is continued from Figure 4. If a pair of maps satisfy the relationship constrained by a directed inhibitory network, rearrangement of columns and rows in a particular procedure (see text) can yield a matrix consisting of diagonal matrices with background of 1-components.

Figure 6. Construction for a matrix consisting of diagonal matrices with background of 1-components implemented in a machine, from left to right. Each block in right matrix represents a Boolean algebra whose atoms are represented by diagonal elements. A lattice is obtained as an almost disjoint union of Boolean algebras.

Figure 7. Hasse diagram of a lattice obtained from a matrix shown in Figure 5 below. Left two diagrams connected by the symbol "+" show two Boolean algebras obtained from two blocks in a given matrix. A whole matrix corresponding to a given matrix is an almost disjoint union of two Boolean algebras, where the least and the greatest elements are common to both Boolean algebras.

Figure 8. Hasse diagram of a lattice obtained from a matrix shown in Figure 6 above. A matrix consists of five blocks. A whole matrix corresponding to a given matrix is an almost disjoint union of five Boolean algebras. Each Hasse diagram of Boolean algebra corresponds to each block in a given matrix. The least and the greatest elements common to both Boolean algebras are represented by circles.

**Figures:**

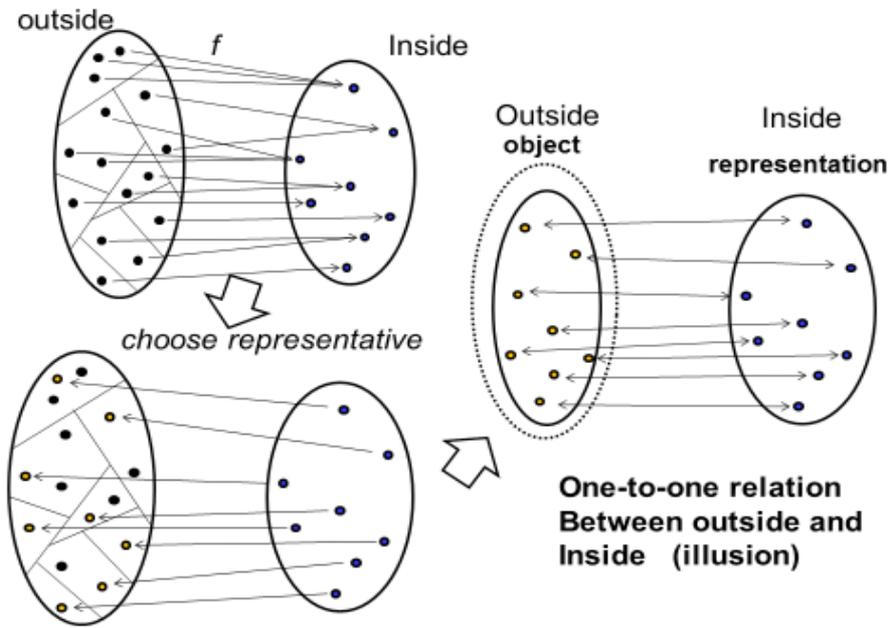

Figure 1

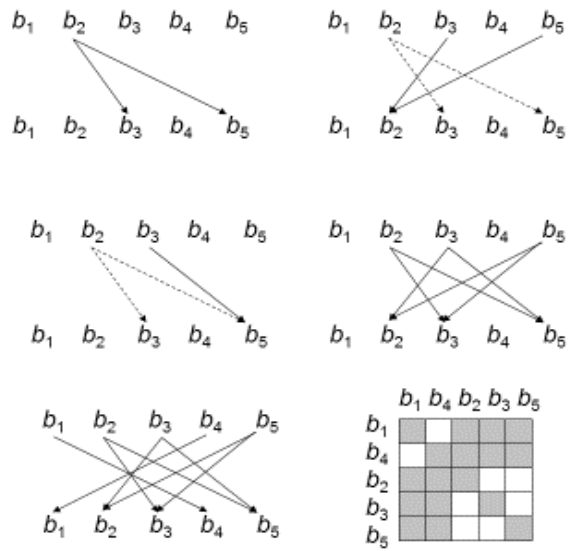

Figure 2

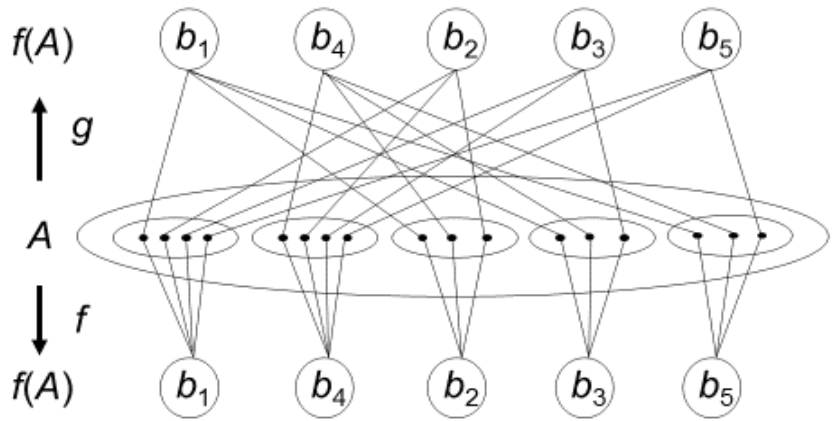

Figure 3

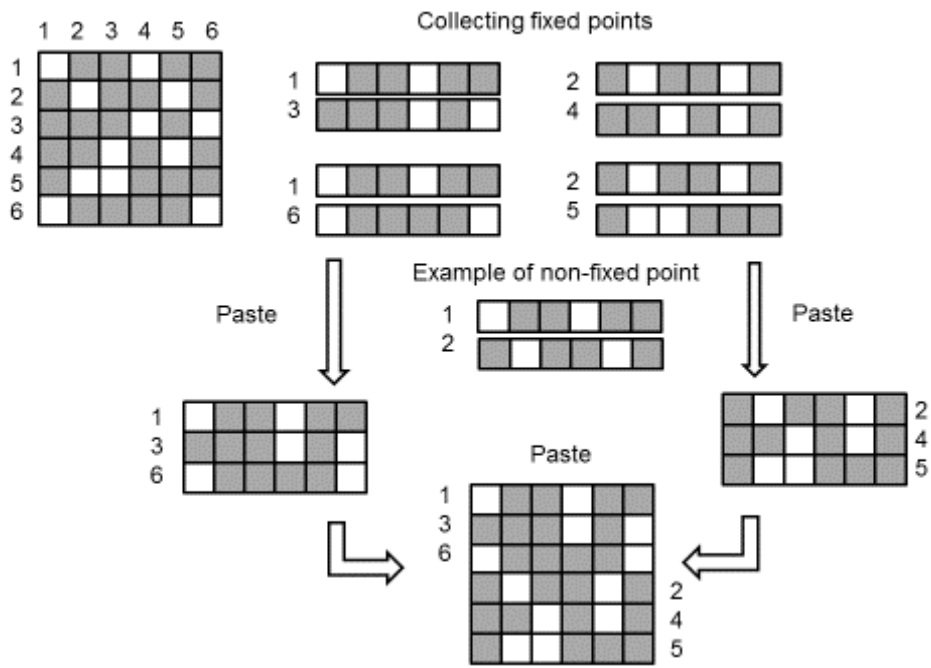

Figure 4

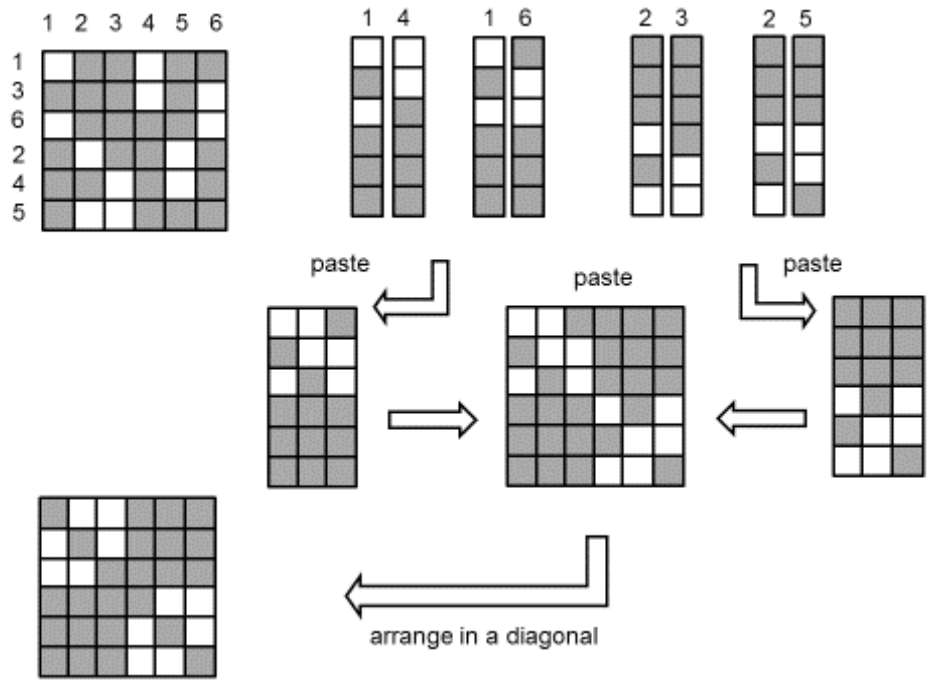

Figure 5

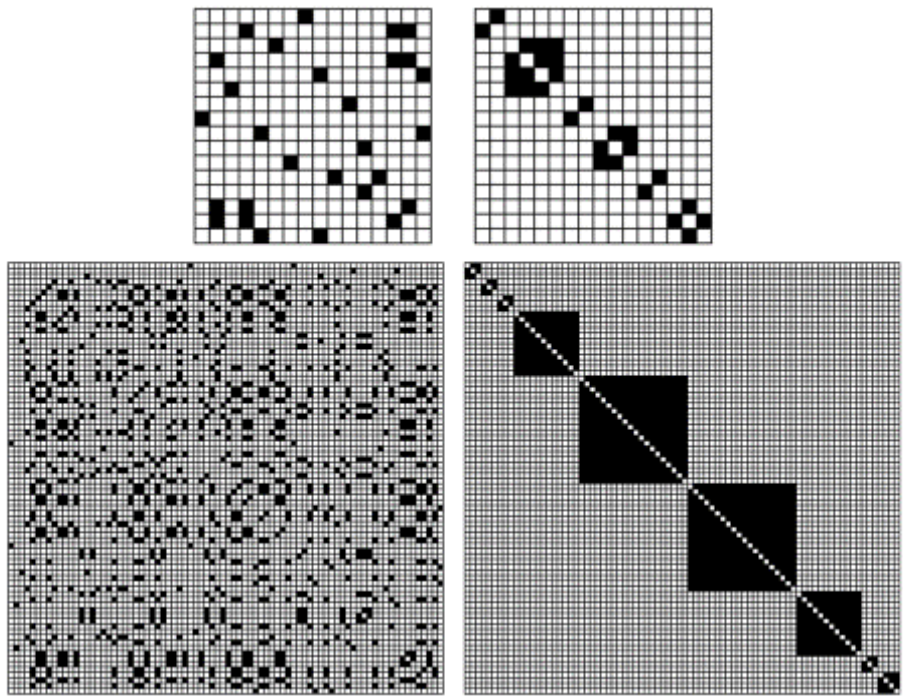

Figure 6

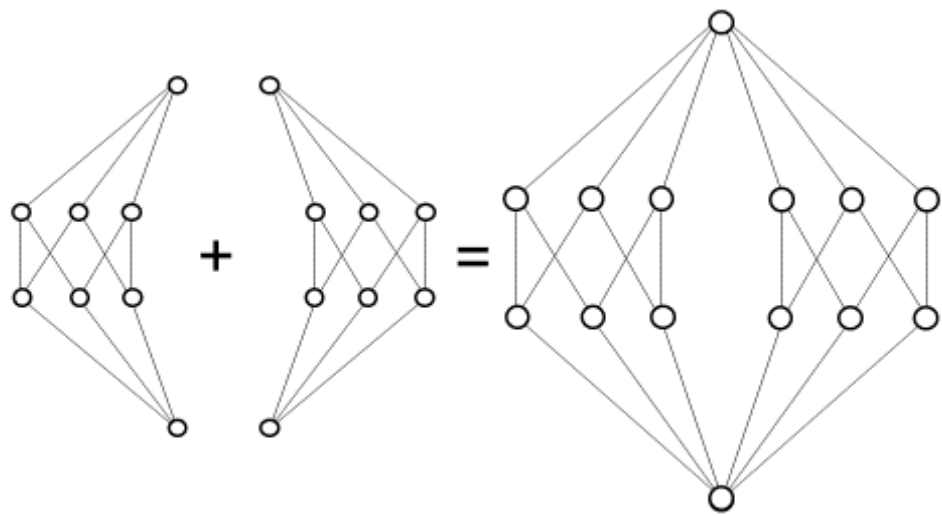

Figure 7.

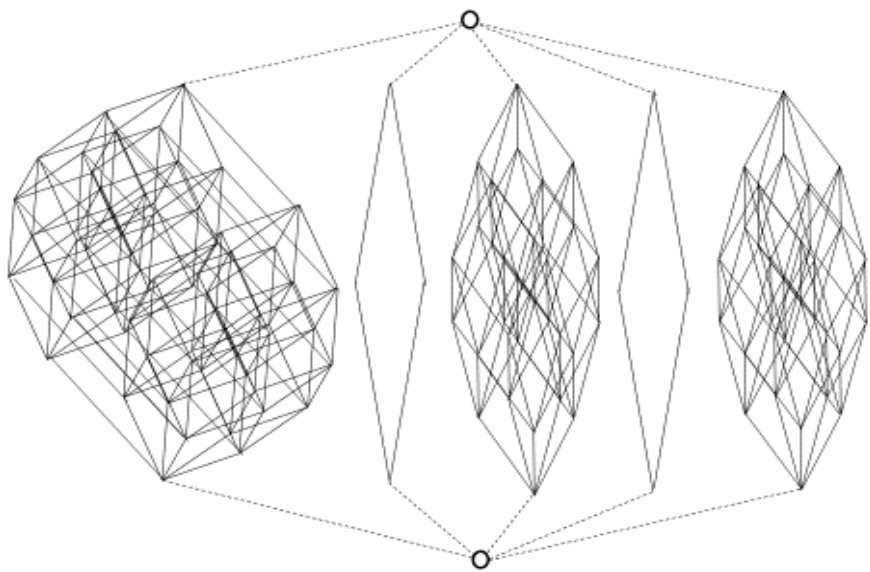

Figure 8